\documentclass[aps,twocolumn,prb,superscriptaddress,floatfix,showpacs]{revtex4}
\usepackage{amsmath}
\usepackage{graphicx}
\usepackage{subfigure}
\usepackage{color}
\usepackage{epstopdf}

\def\lsim{\mathrel{\mathpalette\gl@align<}}
\def\gsim{\mathrel{\mathpalette\gl@align>}}
\def\gl@align#1#2{\lower.6ex\vbox
{\baselineskip\z@skip\lineskip\z@
\ialign{$\m@th#1\hfil##\hfil$\crcr#2\crcr\sim\crcr}}}

\makeatother

\newcommand\ba{\begin{eqnarray}}
\newcommand\ea{\end{eqnarray}}
\newcommand\be{\begin{equation}}
\newcommand\ee{\end{equation}}
\newcommand\bi{\bibitem}
\newcommand{\ct}{\cite}

\def\non{\nonumber}

\def\la{\lambda}

\begin{document}

\title{Tuning the presence of dynamical phase transitions in a generalized $XY$ spin  chain}

\author{Uma Divakaran}
\affiliation{UM-DAE Center for Excellence in Basic Sciences, Mumbai 400 098, India}
\author{Shraddha Sharma}
\affiliation{Department of Physics, Indian Institute of
Technology, Kanpur 208 016, India}

\author{Amit Dutta}
\affiliation{Department of Physics, Indian Institute of
Technology, Kanpur 208 016, India}


\begin{abstract}


We study   an integrable spin chain with three spin interactions and the staggered field  ($\la$) while the  latter  is quenched either
slowly (in a linear
fashion in time ($t$) as $t/\tau$ where $t$ goes from a large negative value to a large positive value and $\tau$ is the inverse
rate of quenching) or suddenly. In the process, the system crosses quantum critical points and gapless phases. We address the question
whether there exist non-analyticities (known as  dynamical phase transitions (DPTs)) in the subsequent real time evolution of the state (reached following the quench) governed by the final time-independent Hamiltonian. In the case of {sufficiently} slow quenching ({when $\tau$ exceeds a critical value $\tau_1$}), we show that DPTs, of the form similar to those occurring for quenching across an isolated critical point, can occur even when the system is slowly driven across more than one critical point and  gapless phases.  More interestingly, in the anisotropic situation we show that DPTs can completely disappear for some values of the anisotropy term ($\gamma$) and $\tau$, thereby establishing {the existence of  boundaries in the $(\gamma-\tau)$ plane between the DPT and no-DPT regions in both
isotropic and anisotropic cases}. Our study therefore leads to a unique  situation when DPTs may not occur even when an integrable model is slowly ramped across a QCP. \       {On the other hand, considering sudden quenches from an initial
value $\la_i$ to a final value $\la_f$, we show that the condition for the presence of DPTs is governed by relations involving 
$\la_i$, $\la_f$ and $\gamma$ and the spin chain must be swept across $\la=0$ for DPTs to occur.} 
 
\end{abstract}


\maketitle


\section{Introduction}

Inspired by the concept of non-analyticities associated with the free-energy density of  a classical system 
at a finite temperature transition marked  by the zeroes of the partition function in a complex temperature plane  \ct{fisher65}  (see also [\onlinecite{lee52, saarloos84}]), recently there has been a proposal of quantum dynamical phase transitions (DPTs) in
a quenched quantum many-body system \ct{heyl13}. Associated non-analyticities are quantified in terms of  the 
overlap amplitude or the Loschmidt overlap (LO) defined for the quenched quantum system. Focussing on the sudden
quenching case and denoting  the ground state of the initial Hamiltonian as  $|\psi_0\rangle$, the Loschmidt overlap is defined as  $G(t)=\langle\psi_0|e^{-iH_ft}|\psi_0\rangle$; here, $H_f$ is the final Hamiltonian  reached after the quenching process. DPTs occur
when the initial state is orthogonal to the evolved state and the LO vanishes.
Generalizing $G(t)$  to 
 $G(z)$ defined in the  complex  time ($z$) plane, one introduces the corresponding dynamical free energy density, $f(z)=-\lim_{L\to \infty}       \log{G(z)}/L^d$, where $L$ is the linear dimension of a $d$-dimensional system. 
One  then looks  for the zeros of the $G(z)$ (or non-analyticities in $f(z)$), known as Fisher zeros.  For a transverse Ising chain, it has been observed \ct{heyl13}  that  when the system is suddenly quenched across the quantum critical point (QCP) \ct{Sachdev,suzuki13}, the lines of  Fisher zeros  cross the imaginary time axis  at  instants of real time  $t^{*}$; at these instants  {the rate function of the return probability} defined as $I(t) = -       \log |G(t)|^2/L$ shows sharp non-analyticities
signaling the occurrence of DPTs.  

The initial observation by  Heyl {\it{et al.}} \ct{heyl13} that DPTs are associated with the sudden quenches across the QCP 
 has been verified in several  studies \ct{karrasch13,kriel14,andraschko14,canovi14,palami15}.
However, subsequently   it has been shown that DPTs 
are not necessarily connected with the passage through the equilibrium QCP and may occur following a sudden quench  even within the same phase (i.e., not crossing the QCP) for both integrable \ct{vajna14} as well as non-integrable models \ct{sharma15}.
Subsequently,  these studies have been generalized to two-dimensional systems \ct{vajna15,schmitt15}  and the role of topology \ct{vajna15} and the  dynamical topological order parameter have been investigated \ct{budich15}.
We note in the passing that the rate function $I(t)$  is related to the Loschmidt echo which has been
studied in the context of decoherence \ct{quan06,rossini07,cucchietti07,venuti10,sharma12,nag12,mukherjee12,dora13,sharma14,suzuki16} and the work-statistics \ct{gambassi11,russomanno15}.
The finite temperature counterpart of the Loschmidt echo \ct{zanardi07_echo}, namely the characteristic function has also been  useful in studies   of  the entropy generation and emergent thermodynamics in quenched quantum systems \ct{dorner12,sharma15_PRE}. {In fact, the rate function (of the return probability) discussed above in the context of DPTs
can be connected to  the work distribution function corresponding to the zero work  in a double quenching experiment\ct{heyl13}.}

%

The periodic occurrences of non-analyticities in the rate function for an integrable model 
was first reported  in
the context of a slow quenching of the transverse field in the
 transverse Ising chain across its QCPs 
\ct{pollmann10}. Very recently,  associated  DPTs have also been  related  to  Fisher zeros 
 crossing the imaginary axis of  the complex time plane \ct{sharma16}.  This is believed to be in general true for
 an integrable model reducible to decoupled two level problems
quenched slowly across its QCP.

In this paper, we extend the previous studies  further to the slow as well as 
sudden quenching of an integrable quantum Ising model with complicated 
interactions across the QCPs (and also gapless phases) and establish that DPTs may 
completely disappear in some situations depending on the quenching rate (or amplitude in sudden quench) and system parameters.
This is an observation that, to the best of our knowledge,  has not been reported earlier particularly for the slow quenching. 
We note at the outset that for the slow quenching, the final state 
is prepared through  the variation of a parameter of the Hamiltonian as  $t/\tau$  
across the QCP  to the final value of time (and hence, of the parameter); 
 on the contrary, for the sudden quenches the final state happens to be the ground state of the initial
Hamiltonian.
In both the cases $G(t)$  describes the subsequent temporal evolution
of the system with the final time-independent Hamiltonian  setting the origin of time ($t=0$) immediately after the quenching
is complete. Let us also note that the numerical calculations are performed for  a finite system, hence Fisher zeros  do not coalesce into a line, rather constitute a set of closely spaced points.

We would also like to mention  that the  slow quenching dynamics across or to a  QCP has been studied in the context of  possible Kibble-Zurek (KZ) scaling \ct{Kibble1976,Zurek1985} of the defect density and the residual energy  \ct{Zurek2005,Polkovnikov2005} which  have
been explored in various situations \ct{dziarmaga05,damski05,cherng06,mukherjee2007,bib:Pellegrini,sengupta2008,deng08,dutta2010,degrandi10,thakrathi12,
canovi141}.
(For reviews, see  [\onlinecite{PolkovnikovRev,dziarmaga10,dutta15}].)

The paper is organized in the following manner:
In Sec. \ref{sec_slow_dpt}, we introduce the connection between Fisher zeros, DPTs   and the slow (as well as sudden) quenching of a generic two-level integrable
model.
In Sec. \ref{sec_generalized}, on the other hand, we introduce a specific model, namely, a  generalized transverse Ising chain with
three-spin interactions and a staggered magnetic field ($\la$) and  present its phase diagram. In Sec. \ref{sec_generalized_dpt} we
focus on slow quenches and  show that DPTs always occur in the isotropic situation
even when the system is quenched across two critical points and gapless phases {if the quenching is not too rapid}. On the contrary, in the anisotropic
case, there is a clear boundary separating the DPT and the no-DPT region; this establishes that 
the slow quenching of an integrable model across its QCP does not necessarily lead to DPTs.      
{Finally, in Sec. \ref{sec_sudden},
 we consider
the sudden quenching of the staggered field and show how the presence of DPTs following the quench is dictated by
relations involving the initial and final values of the field and the anisotropy parameter; it is worth mentioning that the spin chain
must be quenched across $\la=0$ for DPTs to occur.}

\section{Quenches of an integrable model and DPT}
\label{sec_slow_dpt}

Let us consider an integrable model reducible to a two level system for each momenta mode; the system
 is initially ($t \to-\infty$) in the ground
state  {$|1^i_k\rangle$  of the initial Hamiltonian} for each mode. 
We first consider the slow quenching case. The Hamiltonian is characterized by a  parameter $\la$ which
is quenched from an initial value $\la_i$  following the quenching protocol $\la(t) = t/\tau$ to a final value $\la_f$ so chosen 
 that the system crosses the QCP at $\la = \la_c$ in the process. Since the condition for an adiabatic dynamics breaks
 in the vicinity of the QCP, one arrives at a final state (for the $k$-th mode) given by  $|\psi_{f_k} \rangle = v_k |1_k^f\rangle  + u_k |2_k^f\rangle$, with $|u_k|^2 + |v_k|^2 =1$; here, $|1_k^f\rangle$ and $|2_k^f\rangle$ are the ground
state and the excited states of the final Hamiltonian $H_{f_k}(\la_f)$ with corresponding energy eigenvalues {$\epsilon_{k,1}^f$ and $\epsilon_{k,2}^f$}, respectively.  
One can define the LO  for the mode $k$ as  $\langle \psi_{f_k} |\exp(-H_{f_k} z)| \psi_{f_k} \rangle$ and the corresponding dynamical free energy \cite{heyl13},  $f_k(z) = -       \log \langle \psi_{f_k} |\exp(-H_{f_k} z)| \psi_{f_k} \rangle/L$, where $z$ is the
complex time {with $z=R+it$,  $R$ being the real part and $t$ the imaginary part.}
 Summing over the contributions from all the momenta modes and converting
summation to the integral in the thermodynamic limit,  one gets
{
\ba
f(z) &=& -  \int_0^{\pi} \frac{dk}{2\pi}       \log \left( |v_k|^2  \exp(-\epsilon_{k,1}^f z) + |u_k|^2 \exp(-\epsilon_{k,2}^f z) \right) \nonumber \\
&=& - \int_0^{\pi} \frac{dk}{2\pi}       \log \left( (1-p_k) \exp(-\epsilon_{k,1}^f z) + p_k \exp(-\epsilon_{k,2}^f z) \right);
\nonumber\\
\label{eq_free_energy}
\ea
}we reiterate that $t$ is measured from the instant  the final state $|\psi_{f_k}\rangle$ is reached
after the slow quench.

We then immediately find the  zeros (i.e., the Fisher zeros) of the ``effective" partition  function where $f(z)$ is non-analytic
as 
{
\be
z_n(k) = \frac 1 { (\epsilon_{k,2}^f -\epsilon_{k,1}^f) } \left(       \log (\frac { p_k}{1-p_k}) + i \pi(2n+1)\right),
\label{eq_fisher_zero}
\ee}
where $n=0,\pm 1, \pm 2, \cdots $. {It is to be emphasized that in $z_n(k)$, the difference of the eigenvalues $(\epsilon_{k,2}^f -\epsilon_{k,1}^f)$, rather than the eigenvalues
themselves, appear.} The Fisher zeros constitute a line (more precisely,  closely spaced points) corresponding to
each $n$ in
the complex $z$ plane \ct{sharma16}. 
The critical mode $k_c$ (for which the gap in the spectrum vanishes at $\la=\la_c$) remains
frozen in the initial state and hence $p_{k=k_c}=1$, while for modes far away from the critical mode $p_k \to 0$; therefore
$z_n(k)$'s goes from $-\infty$  to $\infty$ in the thermodynamic limit as $k$ changes.  From the  continuity argument,
there must exist one  specific value of $k_{*}$ for which $p_{k=k_*}=1/2$ and ${\rm Re} (z_n(k)|_{k=k_*})$ vanishes; 
from \eqref{eq_fisher_zero} we note that
the lines of Fisher zeros cross the imaginary axis for $k_*$. 

The rate function of the return probability in this case can be evaluated
exactly in the form \ct{pollmann10,sharma16}
 {\ba
I(t) &=& -  \frac  {      \log |G(t)|^2}{L} = 2~ {\rm Re} f(z)\non\\
&=&- \int_{0}^{\pi} \frac{dk}{2\pi} \log\left(1 + 4 p_k (p_k-1) \sin^2 \frac{(\epsilon_{k,2}^f-\epsilon_{k,1}^f)}{2} t \right); \non\\
\label{eq_rate_function}
\ea
the non-analyticities in $I(t)$ appear at the values of the real time $t_n^*$s given by
\be
t_n^* =  \frac{\pi} {(\epsilon_{k_*,2}^f-\epsilon_{k_*,1}^f) }  \left(2n+ 1 \right)
 \label{eq_time}
 \ee
derived by setting ${\rm Re} (z_n(k_*))=0$ in Eq.~(\ref{eq_fisher_zero}) as  the argument of the logarithm in Eq.~(\ref{eq_rate_function}) vanishes
for $k=k_{*}$ when $p_{k=k_*}=1/2$. Again, the time instants $t_n^*$ depend on  $(\epsilon_{k_*,2}^f-\epsilon_{k_*,1}^f)$. However, for the case 
$\epsilon_{k,2}^f = -\epsilon_{k,1}^f  = \epsilon_k^f$, Eq. \eqref{eq_time} gets simplified to
\be
 t_n^* =  \frac{\pi} {\epsilon_{k_*}^f}  \left(n+\frac 1 {2}\right)
 \label{eq_time1}
 \ee
}  

{We shall briefly dwell on the case of sudden quenching \ct{heyl13} when the parameter $\la$ is suddenly
changed from an initial value $\la_i$  to a final value $\la_f$; in this case, the final state $|\psi_{k_f}\rangle$ is the initial ground state $|1^i_k\rangle$ (corresponding to $\la_i$) while
the Hamiltonian gets modified to the final Hamiltonian; the LO for the given mode is given by $\langle 1^i_k |\exp(-H_{f_k} z)| 1^i_k \rangle$. Following a similar line of arguments as above, one can show that the dynamical
free energy has a similar form as in Eq.~(\ref{eq_free_energy}) with $|u_k|^2 \to |{\tilde u}_k|^2 ={\tilde p}_k = |\langle 1^i_k|2_k^f\rangle|^2$ and $|v_k|^2 \to |{\tilde v}_k|^2= |\langle 1^i_k|1_k^f\rangle|^2$. Therefore, one finds a similar expression for the rate function
in Eq.~(\ref{eq_rate_function}) (with $p_k \to {\tilde p}_k$) which shows non-analatyicities at the instants of real time
again given by Eq.~(\ref{eq_time}) when ${\tilde p}_{k=k_*}=1/2$. }

\section{Generalized spin model }

\label{sec_generalized}

In this section, we shall consider a generalized spin-1/2 quantum $XY$ chain with a two sublattice structure
in the presence of a three spin interaction ($J_3 >0$) and a staggered field ($h$) described by the Hamiltonian

\begin{eqnarray}
&H& = -h\sum_i (\sigma_{i,1}^z -\sigma_{i,2}^z) - J_1 \sum_i (\sigma^x_{i,1}\sigma_{i,2}^x + \sigma^y_{i,1}\sigma_{i,2}^y) \nonumber\\
&-&J_2 \sum_i (\sigma^x_{i,2}\sigma_{i+1,1}^x + \sigma^y_{i,2}\sigma_{i+1,1}^y)
- J_3 \sum_i (\sigma^x_{i,1} \sigma_{i,2}^z \sigma^x_{i+1,1} \nonumber\\
&+&\sigma^y_{i,1} \sigma_{i,2}^z \sigma^y_{i+1,1}) -J_3 \sum_i (\sigma_{i,2}^x \sigma_{i+1,1}^z \sigma_{i+1,2}^x \nonumber\\
&+& \sigma_{i,2}^y \sigma_{i+1,1}^z \sigma_{i+1,2}^y),
\label{eq_3spin}
\end{eqnarray}
where  $i$ is the site index and the additional subscript $1 (2)$ defines the odd (even) sublattice. The
parameter $J_1$  describes the $XY$ interaction between the spins on sublattice $1$ and $2$
while  $J_2 $ describes the $XY$ interaction between spins on  sublattice  $2$ and $1$ 
such that  $J_1$ is not necessarily equal to $J_2$. In spite of the complicated nature of interactions,
 this spin chain is integrable and  exactly solvable in terms of a pair of Jordan-Wigner fermions \ct{zvyagin06,qptfermi}
defined on even and odd  sublattices as
$
\sigma_{i,1}^{+}=\left[\prod_{j<i}(-\sigma_{j,1}^{z})(-\sigma_{j,2}^{z})\right]a_{i}^{\dagger},$
and $
\sigma_{i,2}^{+}=\left[\prod_{j<i}(-\sigma_{j,1}^{z})(-\sigma_{j,2}^{z})(-\sigma_{i,1}^{z})\right]b_i^{\dagger},
$
where 
$\sigma_{i,1}^{z}=2a_{i}^{\dagger}a_{i}-1$  and $\sigma_{i,2}^{z}=2b_{i}^{\dagger}b_{i}-1$.
 The Fermion operators $a_{i}$ and $b_{i}$ can be shown to satisfy fermionic anticommutation relations.

\begin{figure}
\includegraphics[height=1.2in]{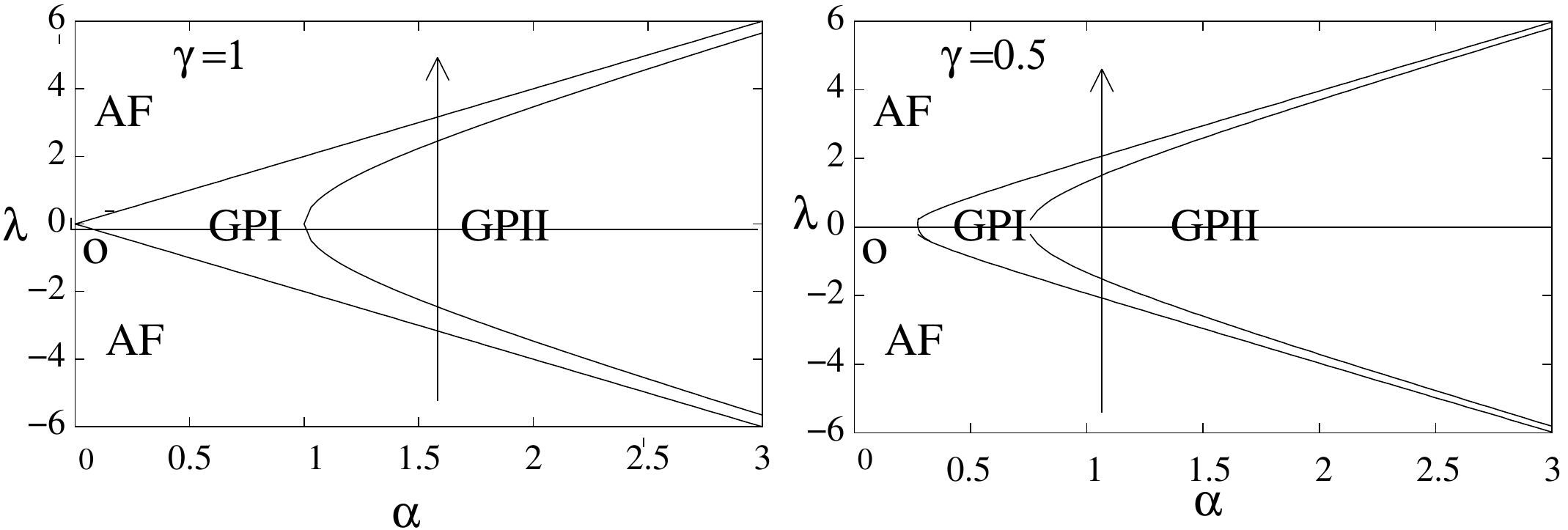} 
\caption{(Color online) Phase diagram of the Hamiltonian \eqref{kit_mat} in the $\alpha-\lambda$ plane for  the isotropic case $\gamma =1$ and 
an anisotropic case with $\gamma=0.5$. For
a fixed $\alpha$ and a 
large magnitude of the rescaled field $\la$, the spin chain is anti-ferromagnetic (AF). On the other hand, when $\la$ is reduced, the system
undergoes quantum phase transitions from Gapless-I (GPI) phase characterized by two Fermi points to the Gapless-II (GPII) phase
characterized by four Fermi points. The vertical line shows the direction of quenching. After [\onlinecite{chowdhury10}].}
\label{fig_gen_spin}
\textbf{}\end{figure}

In the $k$-space, the reduced Hamiltonian is given by
\begin{eqnarray}
H_k= \alpha  \cos k \hat 1 - \frac 1 {2}\left[ \begin{array}{cc} \lambda & -(1+ \gamma e^{-ik}) \non\\
-(1+ \gamma e^{+ik}) & -\lambda \end{array} \right],\\
& & \label{kit_mat} 
\end{eqnarray}
where $\lambda=h/J_1$, $\alpha=J_3/J_1$ and $\gamma=J_2/J_1 $ {and $\hat 1$ is the $2 \times 2$ identity operator and the second part represents the $2 \times 2$ Landau-Zener (LZ) part of the Hamiltonian;} we shall also use the notation $\Delta_k=(1+ \gamma e^{-ik})$
below. {The corresponding eigenvalues of the reduced Hamiltonian $H_k$ are
\ba
{\tilde{\epsilon}}_k^{\pm} &=& \alpha \cos k \pm  \epsilon_k \non\\
&=& \alpha \cos k \pm  \frac 1{2}\sqrt{\la^2 + \gamma^2 +1 + 2 \gamma \cos k}.
\label{eq_eigenvalue}
\ea}
 The  phase diagram obtained by analyzying the spectrum in Eq.~\eqref{eq_eigenvalue}  is shown in  Fig.~\ref{fig_gen_spin}. We shall
 consider the slow as well as sudden quenching dynamics of the Hamiltonian \eqref{eq_3spin} {by varying the parameter $\la$} across the QCPs and gapless phases and probe the corresponding
 DPT scenario. 
 
 As evident from Eq.~(\ref{kit_mat}), the term $\alpha \cos k$ leads to the rich phase diagram of the model under
 consideration  by introducing gapless phases of different kinds, 
{where the gap in the spectrum vanishes solely due to the presence of $\alpha \cos k$.}  
However, this term does not participate in the {dynamics. This is because
 of the fact that the term $\alpha \cos k$ is associated with the identity operator 
which always commutes with the time evolution operator for any type of temporal evolution.}
The dynamics of the system is, therefore, entirely determined
 by the LZ part of Eq.~ (\ref{kit_mat}). This in fact leads to a conspicuous behavior as far as DPTs are
 concerned as we shall discuss below; {furthermore, only the terms $\pm \epsilon_k$ appearing in the eigenvalues in Eq.~ \eqref{eq_eigenvalue} determine the
 instants at which DPTs occur. This is also clear from Eq. \eqref{eq_time} that  only the difference of eigenvalues plays a role in determining $t_n^*$s and hence, the results will be completely independent of the parameter $\alpha$. Additionally, the eigenfucntions of the Hamiltonian $H_k$ are  also identical to those of the LZ part.}  
 
  We note in the passing that Hamiltonian of the form \eqref{eq_3spin} has been studied
 extensively \cite{suzuki71,perk75} over decades; recently topological aspects of this kind of models have also 
 been explored \ct{zhang15}. 

\begin{figure}
\includegraphics[height=3.0in,angle=-90]{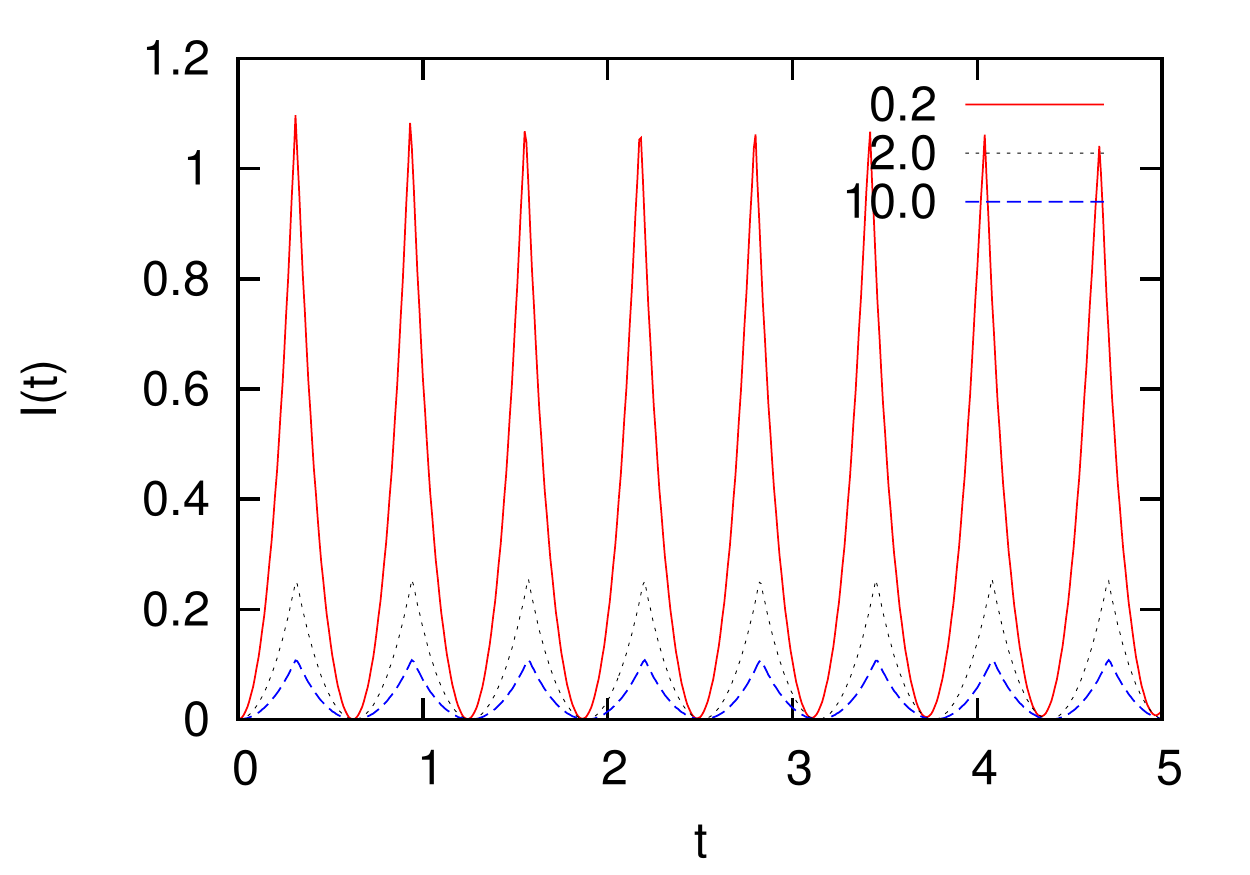}
\includegraphics[height=3.0in,angle=-90]{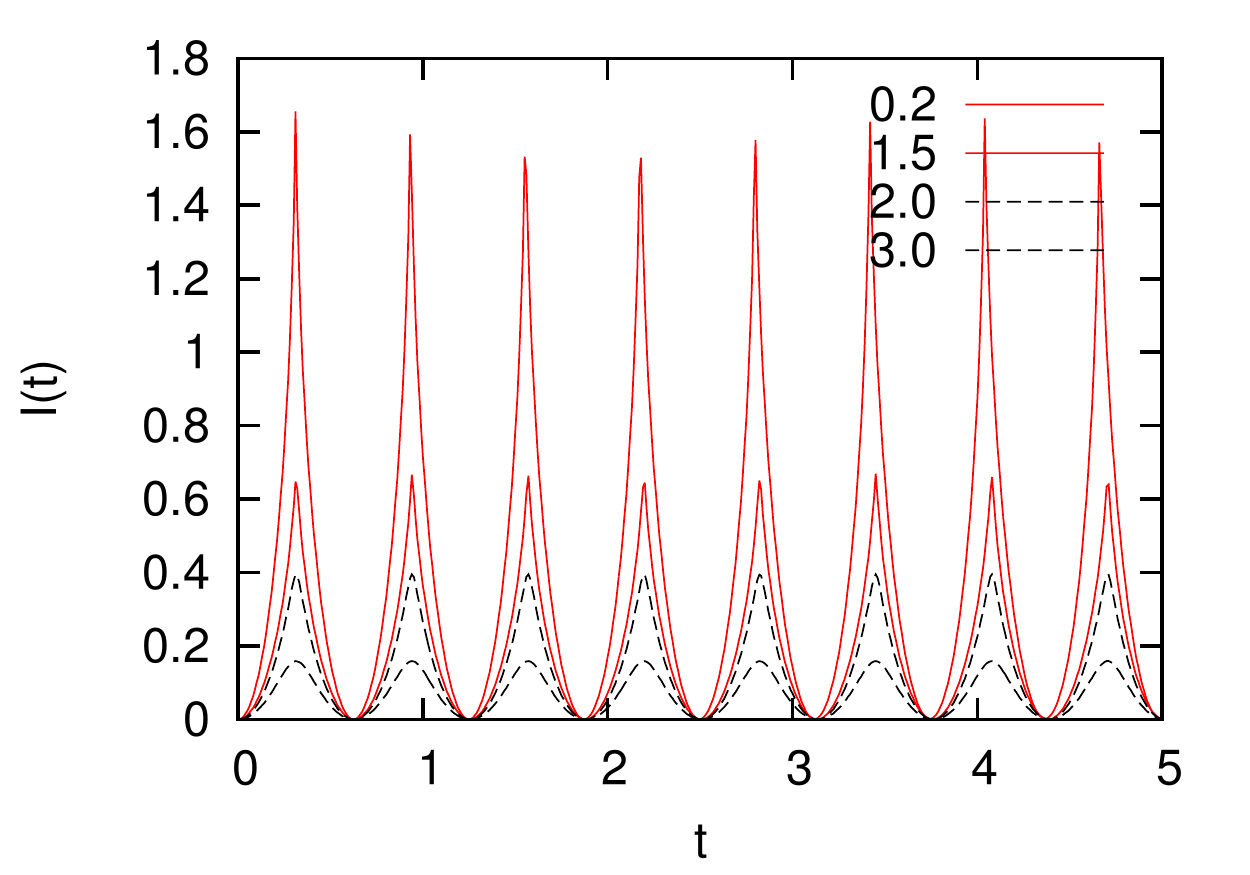}
\caption{(Color online) {The rate function $I(t)$ as obtained by numerically integrating Eq. \eqref{eq_rate_function} is plotted as a function of time following a slow
quench {from a large negative to a large positive value of $\la$.}} The upper panel corresponds to $\gamma=1$ where there are periodic occurrences of  DPTs  for all values of $\tau$'s subject
to the condition that $p_k|_{k=0} <0.5$, i.e., $\tau >\tau_1(\gamma)|_{\gamma \to 1}$.
The lower panel corresponds to the anisotropic situation with $\gamma=0.5$ for which the critical value of $\tau$ as obtained 
from Eq.~(\ref{eq_no_dpt}) is given by $\tau_2(\gamma)=1.76$. As discussed in the text, 
the figure shows the presence of DPTs for $\tau_1(\gamma) <\tau<\tau_2(\gamma)$ while they disappear for $\tau > \tau_2(\gamma)$. The red
colored (solid line) show sharp peaks whereas black colored (dashed lines) show peaks those are rounded off.}
\label{fig_3spin}
\end{figure}

\section{Slow Quenches: DPT-no DPT boundary}

\label{sec_generalized_dpt}

Let us consider a variation of  the field $\lambda =t/\tau$ from $-10$ to $+10$ so that the system is quenched from one antiferromagnetic phase to the other crossing both the gapless phases.    
The probability of excitations $p_k$ following the quench  is  given by the LZ transition probability \ct{landau,sei} $p_k=e^{-\frac{\pi |\Delta_k|^2 \tau}2}$; here, $|\Delta_k|^2 = (1 + \gamma^2 + 2 \gamma \cos k)$ which  vanishes for $k=\pi$ at the boundary
between the anti-ferromagnetic (AF) phase and the gapless phase GPI for  the isotropic case $\gamma=1$. 
{Probing the rate function, we indeed find periodic
occurrences of sharp non-analyticities
as expected in the case of quenching across an isolated QCP 
(see the numerical result presented in the top panel of Fig.~(\ref{fig_3spin})); 
the instants at which these non-analyticities appear can be matched with those obtained 
from Eq.~(\ref{eq_time1}) 
with $\epsilon_{k_*}^f=(1/2)\sqrt{\lambda_f^2+1+\gamma^2+2\gamma\cos{k_*}}=(1/2)\sqrt{\lambda_f^2+2+2\cos{k_*}}$ for $\gamma=1$,  
$\lambda_f$ being the final parameter value reached after the quenching.}  
It is worth mentioning that the dynamics is completely insensitive to the fact that the system
is driven across  gapless phases in the process of quenching and hence no trace of gapless phases is  reflected in DPTs. 
{ However, the occurrences of DPTs
also require the condition that the minimum value of the non-adiabatic transition probability 
$p_{k=0}= \exp({-\frac{\pi |\Delta_k|^2|_{k=0} \tau}2})= \exp({-\frac{\pi (1+\gamma)^2 \tau}2})$ 
must be less than 1/2 so that a $k_*$ (for which $p_{k=k_*}=1/2$) exists. This does not
happen if  the quenching is too rapid, i.e., $\tau < \tau_1(\gamma)=2 \log 2/\{\pi(1+ \gamma)^2\}$ 
for $\gamma \neq 1$; for $\gamma=1$, $\tau_1(\gamma)|{_{\gamma\to 1}}=\log 2/(2 \pi)$. 
One therefore  does not indeed observe DPTs even in the isotropic case for too rapid quenching processes.}

\begin{figure}
\includegraphics[height=2.7in]{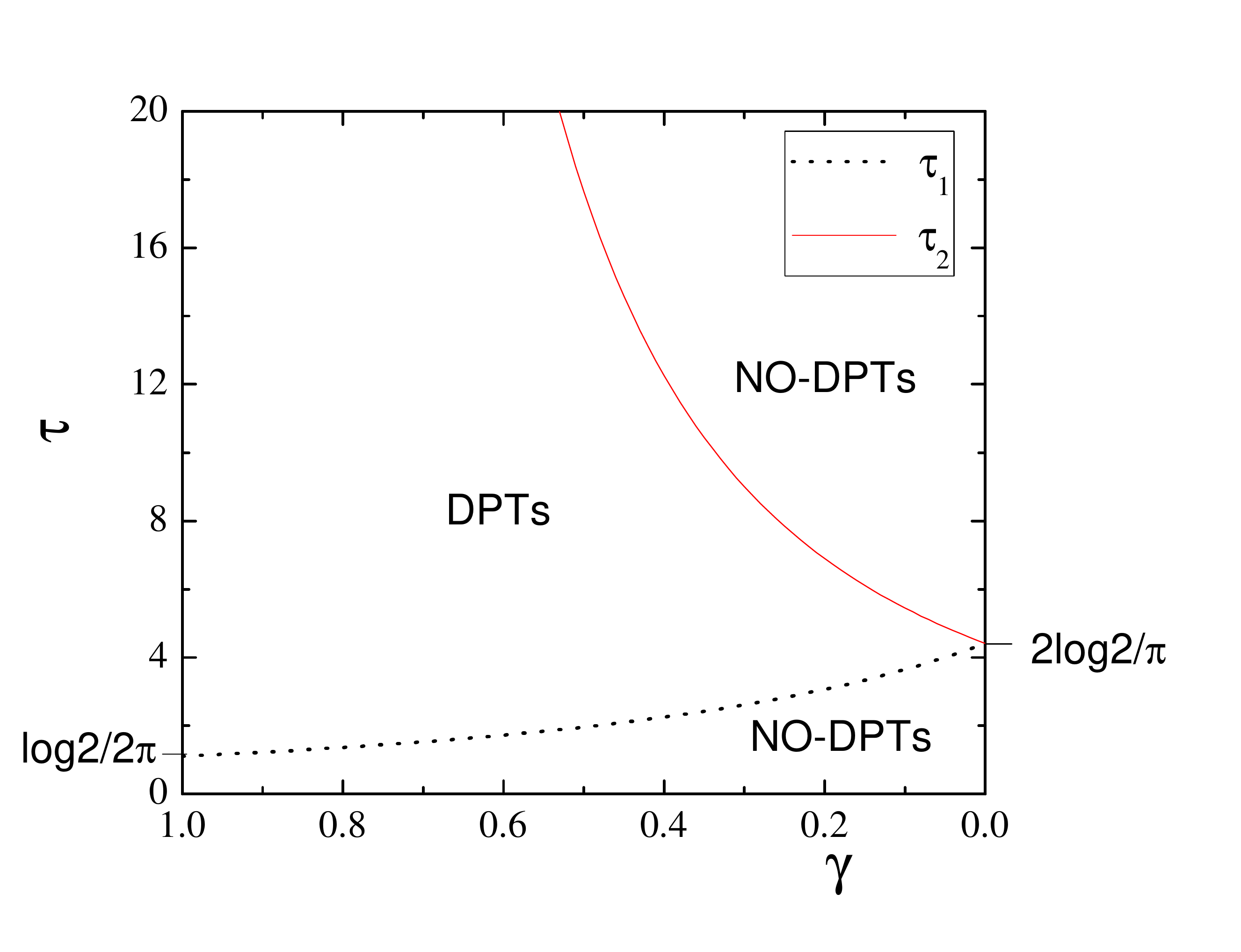}
\caption{(Color online) The phase diagram in the $\gamma$-$\tau$ plane showing the DPT and the no-DPT regions following a slow quench {as
discussed in Fig.~\ref{fig_3spin}}. The upper curve 
corresponds to the condition presented in Eq.~(\ref{eq_no_dpt}) i.e., $\tau_2(\gamma)=2 \log 2/(\pi(1-\gamma)^2)$ which diverges in the isotropic case ($\gamma=1$). The  lower curve denoted by $\tau_1(\gamma)=2 \log 2/(\pi(1+\gamma)^2)$ is obtained from
the requirement $p_{k=0} =1/2$ and $\tau_1(\gamma)|_{\gamma=1}= \log2 /2 \pi$.  It is to be noted that the values of both $\tau_1(\gamma)$ and $\tau_2(\gamma)$ are zoomed by a factor of $10$ for better visibility.  The DPTs exist for the $\gamma$ and $\tau$ values lying in the region bounded by  $\tau_2(\gamma)$ and $\tau_1(\gamma)$.}
\label{fig_slow_phase}
\end{figure}

We now move to the more interesting situation which arises in the anisotropic case ($\gamma \neq 1$); in this case, $|\Delta_k|^2 = (1 + \gamma^2 + 2 \gamma \cos k)$, assumes the minimum value at the boundary between AF and the GPI phase for the mode
$k=\pi$ and is given by $|\Delta_k|^2 = (1-\gamma)^2$, and hence the maximum value of the non-adiabatic transition probability
$p_{k}^{\rm max} = \exp (-\pi (1-\gamma)^2 \tau/2)$.   As emphasized before, DPTs can occur only when  $p_k=1/2$.  If the maximum possible value of $p_{k}^{\rm max}$ is less than $1/2$, no DPT can appear even when the system
is quenched across the QCPs and gapless phases. We therefore find a boundary in the $(\gamma-\tau)$ plane  given
by the equation:
\be
\exp (-\pi (1-\gamma)^2 \tau/2)=1/2;~~~\tau=\tau_2(\gamma)=\frac{2       \log 2}{\pi(1-\gamma)^2}.
\label{eq_no_dpt}
\ee
For a fixed $\gamma$, if $\tau$ exceeds $\tau_2(\gamma) $, DPTs disappear. This is verified numerically and shown
in the lower panel of Fig.~(\ref{fig_3spin}) {where we evaluate $I(t)$ by numerically calculating $p_k$s and using the values $\epsilon_{1,k}^f= -\epsilon_k^f$ and $\epsilon_{2,k}^f= \epsilon_k^f$ (corresponding to $\la_f=10$) in Eq.~\eqref{eq_rate_function};} we show that DPTs occurring for $\tau <\tau_2(\gamma)$, disappear when $\tau$ exceeds 
$\tau_2(\gamma)$. Referring to the situation $\gamma=1$, when maximum value of  $p_k$ (for $k=\pi$)
is equal to unity and $\tau_2(\gamma)|_{\gamma\to 1} \to \infty$,   DPTs periodically appear for all values of {$\tau >\tau_1(\gamma)$}. On the contrary,
for $\gamma \ne 1$, there always exists a critical $\tau_2(\gamma)$  and  DPTs appear only when {$\tau_1(\gamma)<\tau < \tau_2(\gamma)$}.
We would like to emphasize that these
observations are all appropriately supported by the behavior of the lines of Fisher zeros, e.g., one can verify that the lines of Fisher zeros never
cross the imaginary axis in the no-DPT region in the anisotropic case.   {
All these conditions are summarised in Fig.~(\ref{fig_slow_phase}).}

 Interestingly, the relation (\ref{eq_no_dpt}) does not
depend on the parameter  $\alpha$ which plays no role in the temporal evolution of the system. {Let us also addresss the question what happens when $\alpha$ is quenched keeping $\la$ and $\gamma$ fixed. In this case, the initial ground state  of the Hamiltonian $H_k$  (which is  also the ground state of the LZ part of the Hamiltonian \eqref{kit_mat}),  only evolves through an overall phase accumulation. Since the LZ part  is unaltered,
there is no non-trivial dynamics  for any mode $k$,  
and hence, no DPT is expected; the evolved state is never orthogonal to the state $|\psi_{f_k}\rangle$.}
 It is also noteworthy that in the anisotropic case, the defect density shows an exponential decay (with $\tau$) as opposed to the standard power-law
KZ scaling \ct{chowdhury10}.

\section{Sudden quenches: Conditions for DPTs}

\label{sec_sudden}

In this section, we shall consider a sudden quenching of the parameter $\la$ from an initial $\la_i$ to a final value $\la_f$.
We address the questions whether DPTs are always present in the subsequent temporal evolution and how does the situation get altered in the anisotropic
case in comparison to the isotropic case. Remarkably, as we shall illustrate below, in this case also
whether DPTs are present or absent depend on some conditions involving $\la$ and $\gamma$ both for $\gamma=1$ and $\neq 1$.
Referring to the Hamiltonian (\ref{kit_mat}), we find that the ground state and the excited state, i.e., the adiabatic basis
states,  for a given $\la$ (say, $\la_i$)
is given by

\ba
|1_k^i\rangle = \cos \frac{ \theta_k}{2} (1, 0)^T  - \sin \frac{ \theta_k}{2} (0, 1)^T\non\\
|2_k^i \rangle =  \sin \frac{ \theta_k}{2} (1, 0)^T  + \cos \frac{ \theta_k}{2} (0, 1)^T,
\label{eq_wave_function}
\ea
{where $\tan \theta_k =- |\Delta_k|/\lambda$ which clearly does not depend on $\alpha$.}
When the field $\la_i$ is suddenly changed to $\la_f$, the excitation probability is given by 
${\tilde p}_k = |{\tilde u}_k|^2= |\langle 1_k^i| |2_k^f \rangle|^2$. As discussed before, the necessary condition
for the presence of a DPT requires ${\tilde p}_k |_{k=k_*}=1/2$. Using Eqs.~(\ref{eq_wave_function}), we immediately
find
\ba
 {\tilde p}_k &=& |{\tilde u}_k|^2= |\langle 1_k (\la_i)| |2_k(\la_f) \rangle|^2= \sin^2 [(\theta_k^{i} - \theta_k^f)]/2\non\\
 &=&  {\frac 1 {2}\biggl[1 -\frac {\la_f \la_i + |\Delta_k|^2} {\sqrt{(\la_i^2 + |\Delta_k|^2)}\sqrt{(\la_f^2 + |\Delta_k|^2)}}\biggr] ;} \label{eq_pk_half}
 \ea
 {it should be noted that  ${\tilde p}_k$ depends on $\la_i$, $\la_f$ and $\gamma$ but never on $\alpha$}. 
 
  To
predict the presence of DPTs, it is sufficient to analyze $|{\tilde u}_{k=0}|^2$ and $|{\tilde u}_{k=\pi}|^2$; the necessary
condition for DPT would then be $|{\tilde u}_{k=\pi}|^2 > 1/2$ and  $|{\tilde u}_{k=0}|^2 <1/2$. (We recall that the
$|\Delta_k|$ is minimum for the mode $k=\pi$ and hence probability of excitation is maximum for that particular mode).
If these conditions are satisfied, from the argument of continuity one concludes that there must exist a $k_*$ for
which  $|{\tilde u}_{k=k_*}|^2 = 1/2$, ensuring the existence of DPTs. {We note that this is the most generic
condition for DPTs to occur as long as one can sharply define a $k_*$.} Using \eqref{eq_pk_half}, one can show that 
 ${\tilde p}_k$ becomes equal to $1/2$ for a mode $k$ only when

 \be
 \frac {\la_f \la_i + |\Delta_k|^2} {\sqrt{(\la_i^2 + |\Delta_k|^2)}\sqrt{(\la_f^2 + |\Delta_k|^2)}}=0 \implies \la_f \la_i + |\Delta_k|^2=0, \label{eq_sudden_cond1}
 \ee
 where $|\Delta_k|^2$ is evaluated at the corresponding value of $k$.
  To illustrate the main point in a transparent manner, we choose $\la_f = -\la_i = \la$, (or the other way round i.e.,  $\la_f = -\la_i = -\la$) for which Eq.~(\ref{eq_sudden_cond1})
 assumes a simpler form:
 \be
 \frac {-\la^2 + |\Delta_k|^2} {\sqrt{(\la^2 + |\Delta_k|^2)}\sqrt{(\la^2 + |\Delta_k|^2)}}=0 \implies \la^2 = |\Delta_k|^2.
 \label{eq_sudden_cond2}
 \ee
 
\begin{figure}
\includegraphics[height=3.0in]{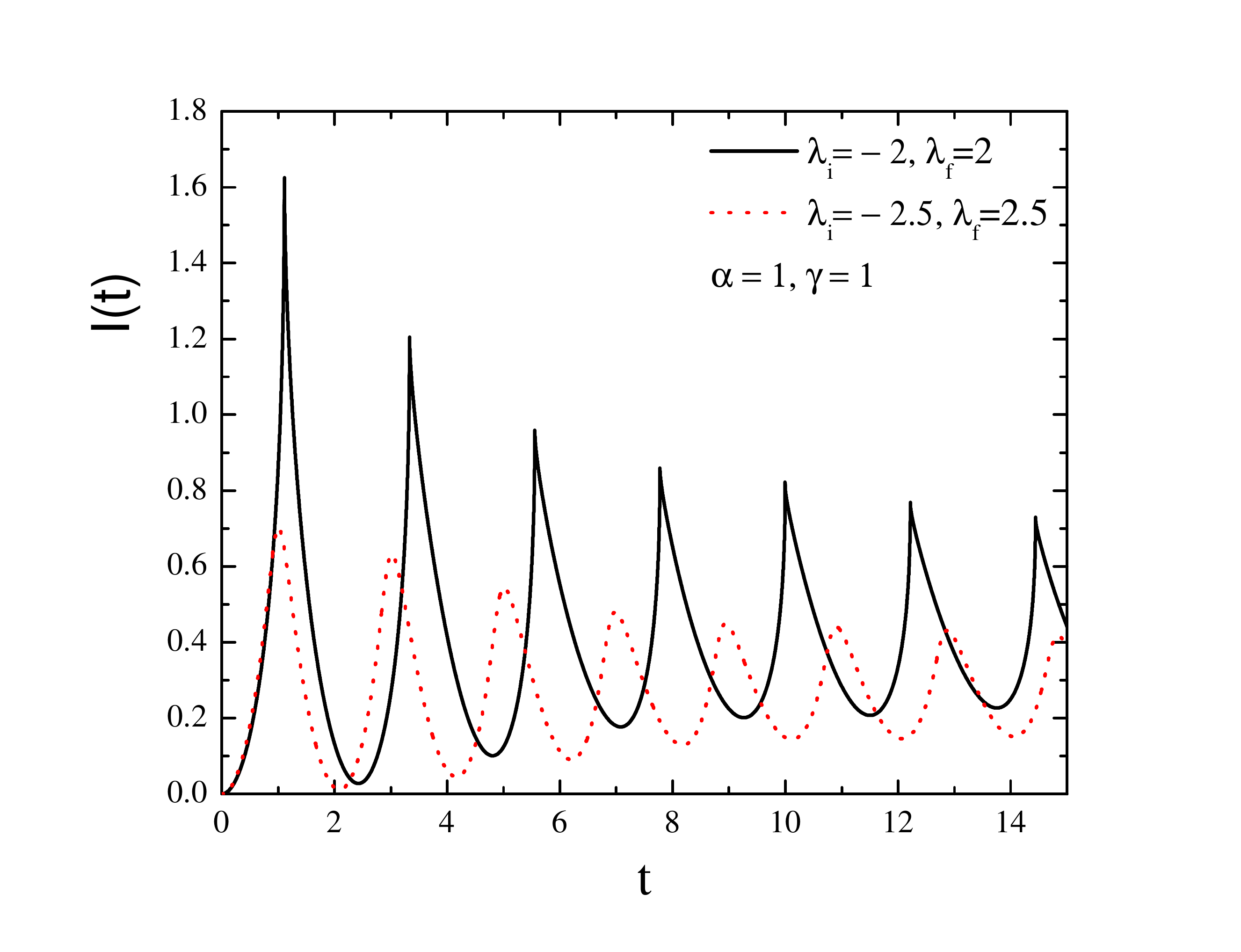}
\includegraphics[height=3.0in]{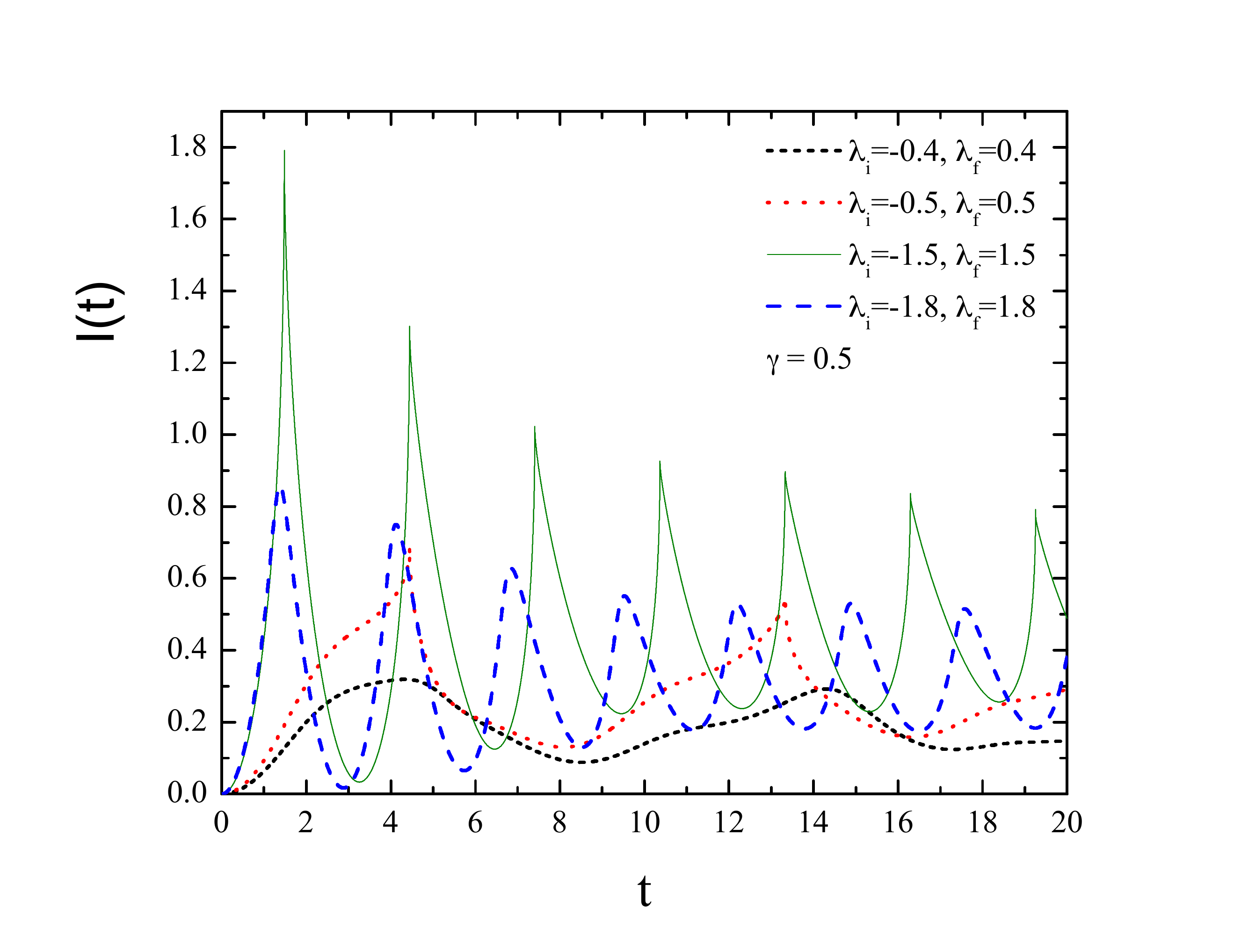}
\caption{(Color online) The presence and the absence of DPTs following a sudden quenching of $\la$ from $-\la$ to $+\la$ for $\alpha=1$.
The upper panel corresponds to $\gamma=1$ where there are periodic occurrences of  DPTs when $\la=2$ while DPTs
get rounded off when $\la=2.5$.
The lower panel corresponds to the anisotropic situation with $\gamma=0.5$ where the presence of DPTs are wiped out
when $\la=1.8$  ($>(1+\gamma)$) or $\la=0.4$ ($<(1-\gamma$)) while these are prominently
present when for $\la=1.5~ (=(1+\gamma))$ and $\la=0.5 ~(=(1-\gamma))$  as discussed in the text. It should also be noted that the position of the maxima
(or non-analyticities) depend on the magnitude  $\la$. }
\label{fig_sudden}
\end{figure}

We shall analyze the condition given in \eqref{eq_sudden_cond2} for the modes $k=0$ and $k=\pi$ for both $\gamma=1$
and $\gamma \neq 1$. In the former case ($\gamma=1$), $|{\tilde u}_{k=\pi}|^2=1$ as the off-diagonal terms of the LZ part of the Hamiltonian
(\ref{kit_mat}) vanish for $k=\pi$ so that this mode is temporally frozen. On the other hand, the
condition that $|\tilde {u}_{k=0}|^2 \leq 1/2$, demands $\la^2 \leq  \Delta_k^2|_{k=0} =4$. This implies that whenever the field
$\la$ is quenched from a value $\la_i \geq -2$ to a final value $\la_f  \leq +2$, DPTs will indeed appear. Otherwise,
they are absent. This is numerically verified as shown in the upper panel of Fig.~\ref{fig_sudden}.

Proceeding to the anisotropic case, we find from Eq.~\eqref{eq_sudden_cond2} that the condition $|{\tilde u}_{k=\pi}|^2 \geq 1/2$, leads to $\la \geq (1-\gamma)$
while the requirement  $|{\tilde u}_{k=0}|^2 \leq 1/2$, yields $\la \leq (1+\gamma)$. Therefore, for a sudden quench from $-\la$ to $+\la$ with
a given $\gamma$, one finds a range of $\lambda$ dictated by the condition $(1-\gamma) \leq \la \leq (1+ \gamma)$ for
which DPTs would appear as numerically verified in the lower panel of Fig.~\ref{fig_sudden}. This condition immediately reduces to the isotropic case for
$\gamma=1$, where, as shown above, the magnitude
of $\lambda$ should be less than 2 to observe DPTs.

\begin{figure}
\includegraphics[height=3.0in]{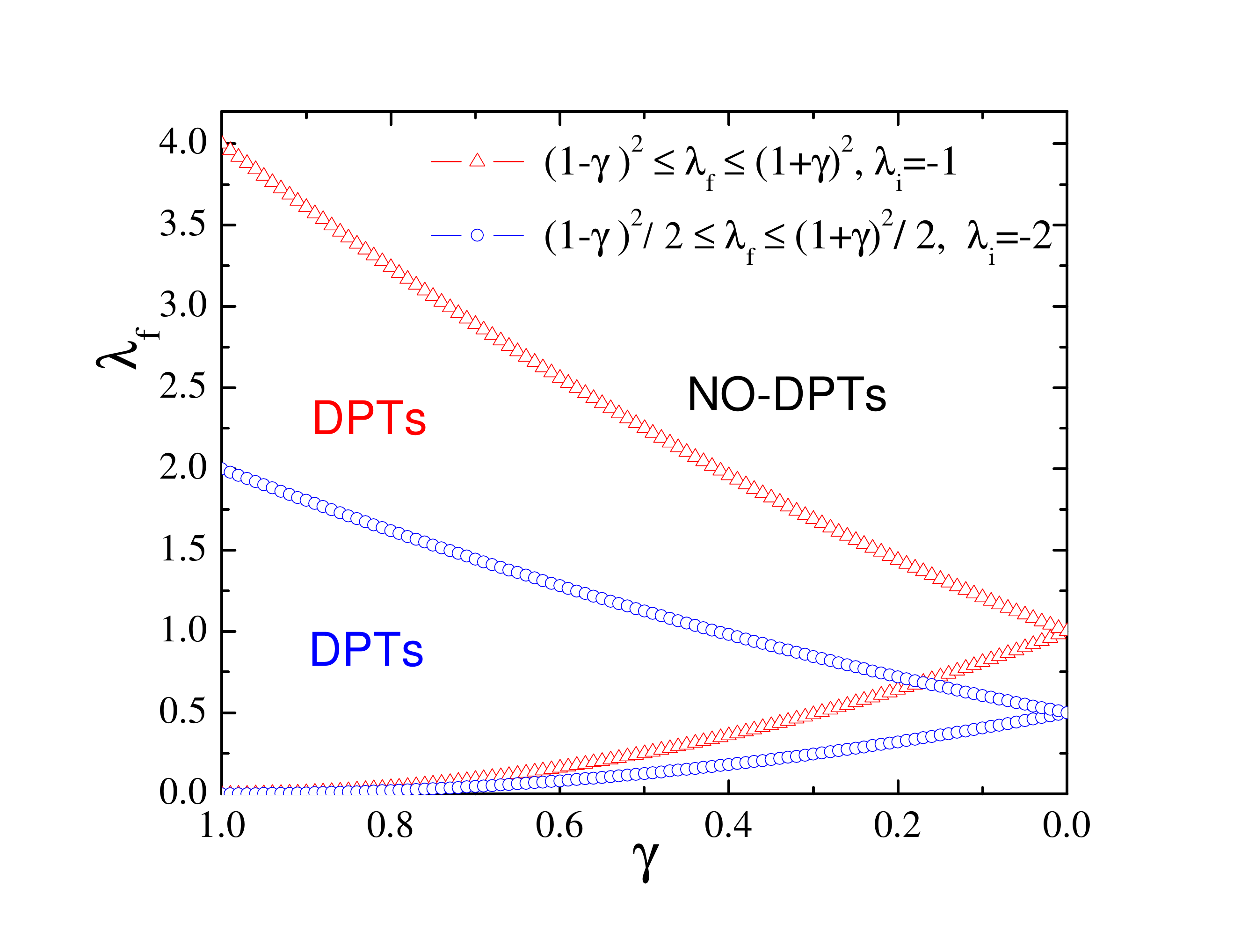}
\caption{(Color online) The phase diagram in the $\la_f-\gamma$ plane showing the regions where DPTs will occur following  sudden quenches with $\la_i=-1$ (red triangles) and $=-2$ (blue
circles) to a final value $\la_f$. In both the cases,  DPTs occur when $\la_f$ lies within the range $(1-\gamma)^2/|\la_i|$ and $(1+\gamma)^2/|\la_i|$ as elaborated in the text.  In the isotropic situation  ($\gamma=1$),  the condition
gets simplified to $0 \leq \la_f \leq 4/|\la_i|$. }
\label{fig_sudden_phase}
\end{figure}

Referring to Eq.~\eqref{eq_sudden_cond1}, we find that for DPTs to occur,  the quantity $\la_i \la_f$ must be
negative; that implies that the spin chain must be quenched across $\la=0$. In that sense, the line $\la=0$ is special;
this is in congruence with the observation reported in the Ref. [\onlinecite{divakaran13}] where it has been shown
 the Loschmidt echo when studied as a function of
$\la$ shows a dip only at $\la=0$, thereby detecting only a special point of the phase diagram. Therefore for a generic
situation, the condition for DPT to occur would be  $|{\tilde u}_{k=\pi}|^2 > 1/2$ and  $|\tilde u_{k=0}|^2 <1/2$
along with the condition the system is   quenched across $\la=0$; for a quench from an initial value $-\la_i$ to
a final value $\la_f$, Eq.~\eqref{eq_sudden_cond1} then leads to a more generic condition $(1-\gamma) \leq \sqrt{|\la_i| \la_f} \leq
(1+\gamma)$ for DPTs to occur. ({If the quenching is from $+\la_i$ to $-\la_f$, the condition gets modified to $(1-\gamma) \leq \sqrt{\la_i |\la_f|} \leq
(1+\gamma)$.)} This has also been numerically verified.  What needs to be emphasized is that whether DPTs 
are present following a sudden
quench is completely independent of the fact whether the system is quenched across a QCP or not; therefore, the passage
through a QCP is never a necessary criteria. All these above  conditions are summarized in Fig.~ \ref{fig_sudden_phase}

 \section{Conclusion}

We have explored the  possibility of DPTs following  slow  as well as sudden  quenches of a model Hamiltonian with a rich phase diagram with two gapless phases. 
We find some worth mentioning 
results not reported before. The term of the reduced Hamiltonian  that results into these gapless phases  do not participate in the dynamics and hence the passage through the gapless phases is not reflected in the behavior of DPTs those may occur following the quench both in isotropic and anisotropic cases.  Consequently, for the slow quenches
in the isotropic case, there are periodic occurrences of DPTs as expected in the case of a slow passage of an integrable model through an isolated QCP {if the quenching
is not too rapid (i.e., for $\tau >\tau_1(\gamma)$)}. On the contrary, in the
anisotropic case, one finds a region in which  {DPTs exist bounded by two limiting quenching rates  $\tau_1(\gamma)$ and $\tau_2(\gamma)$ in the $\gamma-\tau$ plane as summarized in
Fig.~\ref{fig_slow_phase}; in the isotropic case {$\tau_1(\gamma)|_{\gamma \to 1} = 2 \log 2/\pi$ and} $\tau_2(\gamma)|_{\gamma \to 1} \to \infty$}.
This model
provides a unique example of  a situation where DPTs could be absent even
 when an integrable model is slowly ramped across a QCP.
    
    Concerning the sudden quenches we find that even in the isotropic case the presence of DPTs is not
guaranteed; neither the situation is like a sudden quench through a single QCP as in the case of slow quenches. Rather
both in the isotropic and anisotropic cases, one finds restrictions on the values of $\la_i$ and $\la_f$ depending on
the parameter $\gamma$  determined from Eqs.~(\ref{eq_sudden_cond1}) and ~(\ref{eq_sudden_cond2}). It is never
important whether the spin chain is quenched across the QCP in the process of quenching;  
however, it  should necessarily be swept through  $\la=0$, i.e.,  either $\la_i$ or $\la_f$ should be
negative for DPTs to appear. {We have illustrated these different scenarios in Fig.~\ref{fig_sudden_phase}}. This is remarkable that DPTs can be made to appear (or disappear)
in the same model by tuning either the anisotropy term  $\gamma$ or the inverse quench rate $\tau$ for slow quenches,
and $\la_i$ and $\la_f$  for sudden quenches such that  the system must be driven  across $\la=0$.


\begin{acknowledgments}
We acknowledge Jun-ichi Inoue  for discussions and Anatoli Polkovnikov and Sei Suzuki for collaboration in related works. Shraddha Sharma acknowledges
CSIR, India and also DST, India, and AD  and UD acknowledges DST, India,  for financial support. AD and SS acknowledge 
Abdus Salam ICTP for hospitality where the initial part of the work was done.
\end{acknowledgments}

\vspace{-\baselineskip}

\end{document}